\pdfoutput=1	

\documentclass{article}   
\usepackage{graphicx}
\usepackage[margin=4cm]{geometry}
\usepackage{bm}	

\begin{document}

\title{User guide to TIM, a ray-tracing program for forbidden ray optics}


\date{}

\maketitle

\begin{figure}[ht]
\begin{center} \includegraphics[width=\columnwidth]{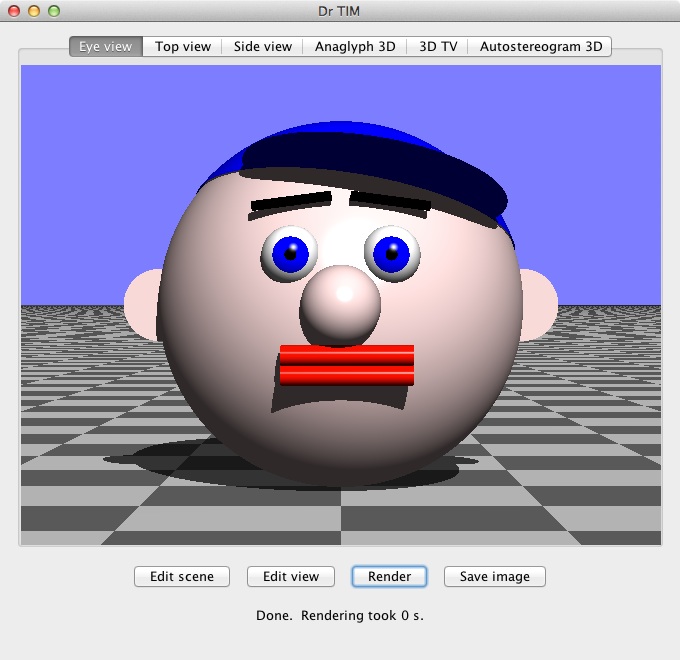} \end{center}
\caption{\label{home-screen-figure}TIM's Java-application window after startup.
This screen shot shows TIM running on a 2.9\,GHz MacBook Pro.}
\end{figure}

\section{Introduction}

\noindent
TIM is a ray-tracing program that can model physically impossible ray optics\footnote{We initially wrote TIM as a tool for our research on windows that perform very general light-ray-direction changes, so general, in fact, that they can lead to physically impossible light-ray fields \cite{Hamilton-Courtial-2009,Courtial-Tyc-2012}.
They can do this by compromising other aspects of the light field, but this compromise can be unnoticeably small.
We call such components METATOYs \cite{Hamilton-Courtial-2009,Wikipedia-METATOY}.
The name TIM started off as an acronym for \textit{\underline{T}he \underline{I}nteractive \underline{M}ETATOY}, but is now beginning to morph into the name of the head with the luscious lips that appears when TIM starts up (Fig.\ \ref{home-screen-figure}).}.
The original version of TIM was described in Ref.\ \cite{Lambert-et-al-2012}.
Since then, a number of specialist intellectual abilities were added to TIM, culminating the award of a PhD to TIM, whose full title is now Dr TIM~\cite{Oxburgh-et-al-2013}.
To friends, however, he is still just ``TIM''.

We built a number of other special features into TIM.
These include the abilities to
\begin{itemize}
\item focus on non-planar surfaces,
\item model relativistic camera speeds,
\item visualise light-ray trajectories,
\item render scenes for 3D viewing.
\end{itemize}
A more detailed explanation of these specialities, and of TIM's more common features, can be found in Refs \cite{Lambert-et-al-2012,Oxburgh-et-al-2013}.

This user guide aims to introduce the use of TIM's interactive user interface. 
Another way of using TIM is described in Ref.\ \cite{Lambert-et-al-2012}, which contains information about using writing code that uses TIM's Java methods, about extending its capabilities, and generally about its underlying code structure and algorithms.
The remainder of this user guide is exclusively about controlling TIM through the interactive user interface, and any mention of ``TIM'' refers to the interactive version.

TIM (in its interactive incarnation) can be run as an applet or as a Java application.
The TIM applet can be embedded in web pages (see \url{http://tinyurl.com/timray}).
The Java application comes packaged as a JAR file, which can be downloaded from TIM's home at the University of Glasgow
(\url{http://tinyurl.com/timray})
or his second home, at sourceforge (\url{https://sourceforge.net/projects/timray/}), and which can be run directly in any good operating system.
The interface of the Java-application version is almost identical to that of the applet version, but additionally allows saving of the calculated images, something the applet version cannot do because of security restrictions designed to protect users from malicious applets.

When it starts up, TIM renders a default scene with default view parameters (position, aperture size, quality, etc.) (Fig.\ \ref{home-screen-figure}).
Section \ref{scene-section} explains how to alter this scene, and then render it.
Section \ref{views-section} outlines TIM's coordinate system, and how to select and interpret the different view positions.
Section \ref{focussing-section} is about focussing in general, and section \ref{non-planar-focussing-section} is specifically about focussing on non-planar surfaces.
Section \ref{light-rays-section} explains how to visualise light-ray trajectories.
Finally, section \ref{parametrisation-section} describes the way scene objects are parametrised in TIM.

TIM requires version 1.6 or higher of the Java Virtual Machine (JVM).

\section{\label{scene-section}Defining the scene and rendering}

\noindent
After TIM has started up, it spends a few moments rendering the default scene and it displays the rendered view (Fig.\ \ref{home-screen-figure}).
In addition to the chequered floor and blue sky, the default scene consists of Tim's distinctive head.
The default scene also contains an eye at the position of the default camera (section \ref{focussing-section}); by default, this eye is invisible.

\begin{figure}
\begin{center} \includegraphics[width=\columnwidth]{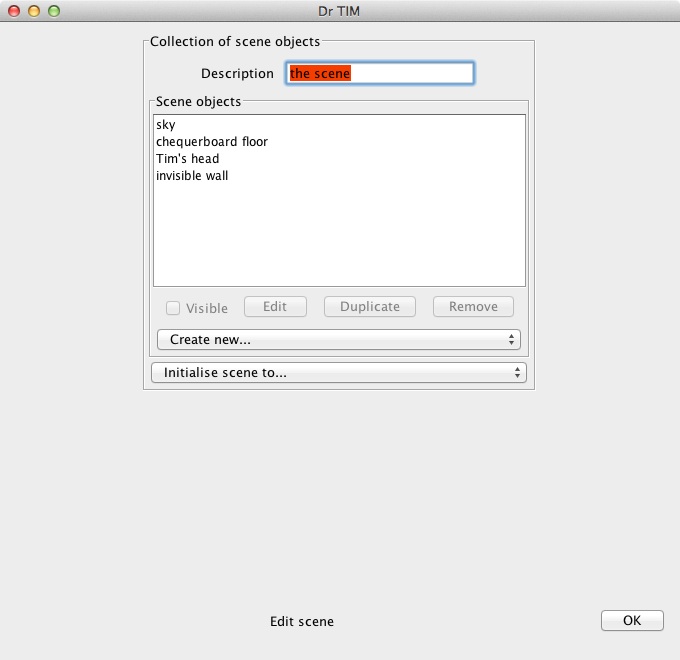} \end{center}
\caption{\label{edit-scene-dialog-figure}``Edit scene'' dialog.
The central list of scene objects contains the scene objects in the default scene.}
\end{figure}

The scene can be altered by clicking on the ``Edit scene'' button.
TIM now shows a list of scene objects (Fig.\ \ref{edit-scene-dialog-figure}).

\begin{figure}
\begin{center} \includegraphics[width=8.4cm]{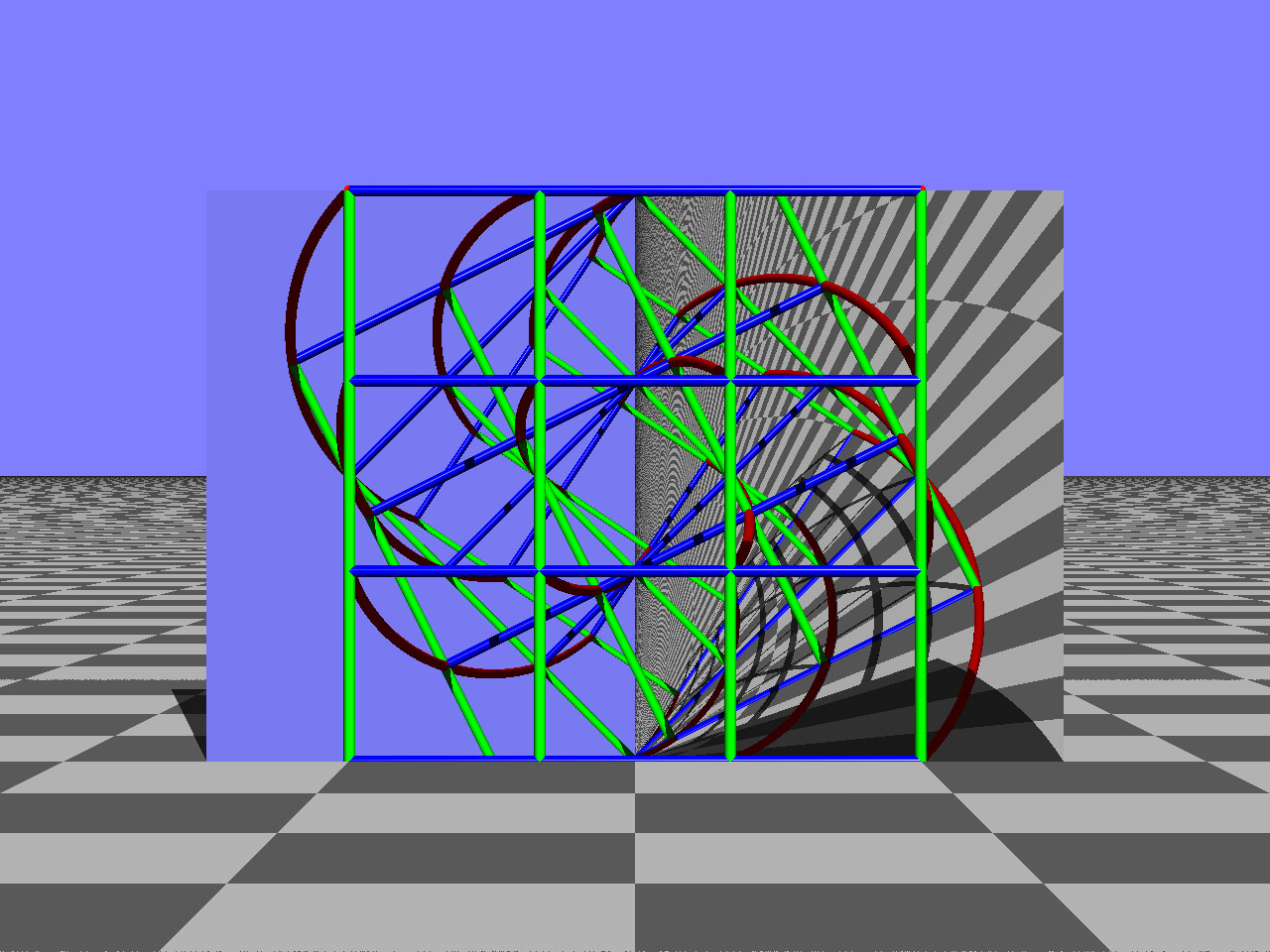} \end{center}
\caption{\label{changed-scene-render-figure}Default lattice behind a window that rotates the light-ray direction through $90^\circ$ around the window normal.}
\end{figure}

As an example for defining a scene in TIM, we alter the scene in a way that allows us to understand better the effect of an example of a METATOY, namely a ray-rotating window \cite{Hamilton-et-al-2009}.
First we remove the head by clicking on ``Tim's head'' in the list of scene objects, and then clicking the ``Remove'' button.
Similarly, we remove the scene object ``invisible wall''.
We add a rectangular window to the scene by selecting ``Rectangle'' in the ``Create new...'' drop-down menu.
This brings up a dialog in which the rectangle's geometrical and optical parameters can be edited; by clicking the ``OK'' button without altering any of the values, we accept the default values of the parameters, which are suitable for our current purposes.
The default optical properties are those of an idealised METATOY that rotates the direction of light rays by $90^\circ$ around the surface normal \cite{Hamilton-et-al-2009}; each light ray emerges from the METATOY with the new direction from the same position where it intersected it, but on the other side of its surface.
Next, we add a suitable three-dimensional lattice behind the window.
We do this by selecting ``Cylinder lattice'' from the ``Create new...'' drop-down menu and then clicking the ``OK'' button at the bottom right of the dialog that appears, thereby accepting the default values of the parameters, which are again suitable for our purposes.
We are now back in the ``Edit scene'' dialog, but the list of scene objects is different from that shown in Fig.\ \ref{edit-scene-dialog-figure}.
By clicking ``OK'', we accept the altered scene.
Now we are back in TIM's main screen, with the status line at the bottom saying ``Ready to render.''
Clicking on the ``Render'' button now renders the altered scene; the result is shown in Fig.\ \ref{changed-scene-render-figure}.

\section{\label{views-section}Coordinates and views}

\noindent
The tabs at the top of the TIM applet area (Fig.\ \ref{home-screen-figure}) allow the selection of different views.
Only one of them, the ``Eye view'', is rendered automatically when TIM starts up.
To render any view, or to re-render it after the scene and/or view parameters have been changed, select the appropriate tab and click on the ``Render'' button.

The standard view is ``Eye view''.
This simulates a camera by default with its aperture centred at coordinates $(0, 0, 0)$, pointing in the positive $z$ direction, and with a horizontal angle of view of $\pm \arctan (2/10) \approx \pm 11.3^\circ$; all these parameters can be varied in Dr TIM (but not in TIM).
The eye-view camera can be either a pinhole camera or a camera with a finite aperture, leading to blurring of out-of-focus objects (section \ref{focussing-section}).
The camera is configured such that greater $x$ values appear further right, and that greater $y$ values appear higher up.
As the $z$ axis points away from the camera, the coordinate system is left-handed.
TIM uses dimensionless coordinates.
Some sense of scale can be gained from the default floor position (in the plane $y=-1$) and patterning (the tiles are squares of side length 1).
The ``Eye view'' can also simulate aspects of moving with relativistic speed through the scene (section \ref{relativistic-section}).

In ``Top view'', everything is seen from directly overhead.
In other words, ``Top view'' shows an orthographic projection into the horizontal, i.e.\ $(x, z)$, plane.
Similarly, in ``Side view'' everything is seen from the positive $x$ direction, so it shows an orthographic projection into the $(z, y)$ plane.

\begin{figure}
\begin{center} \includegraphics{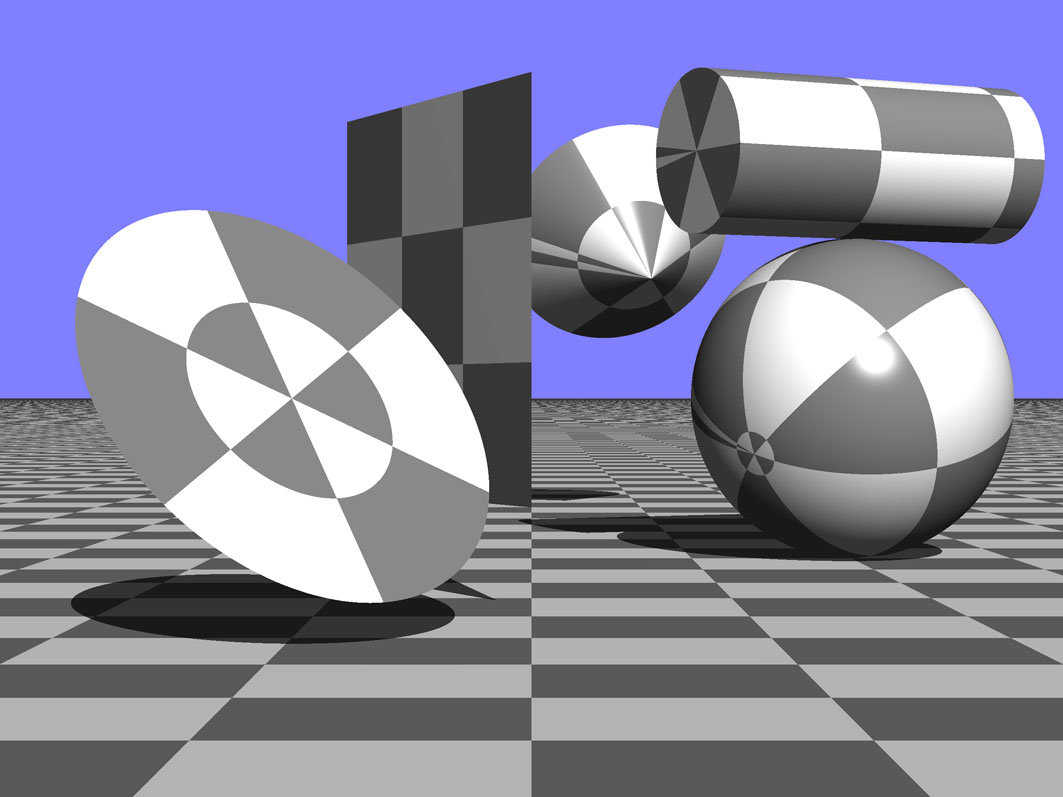} \end{center}
\begin{center} \includegraphics{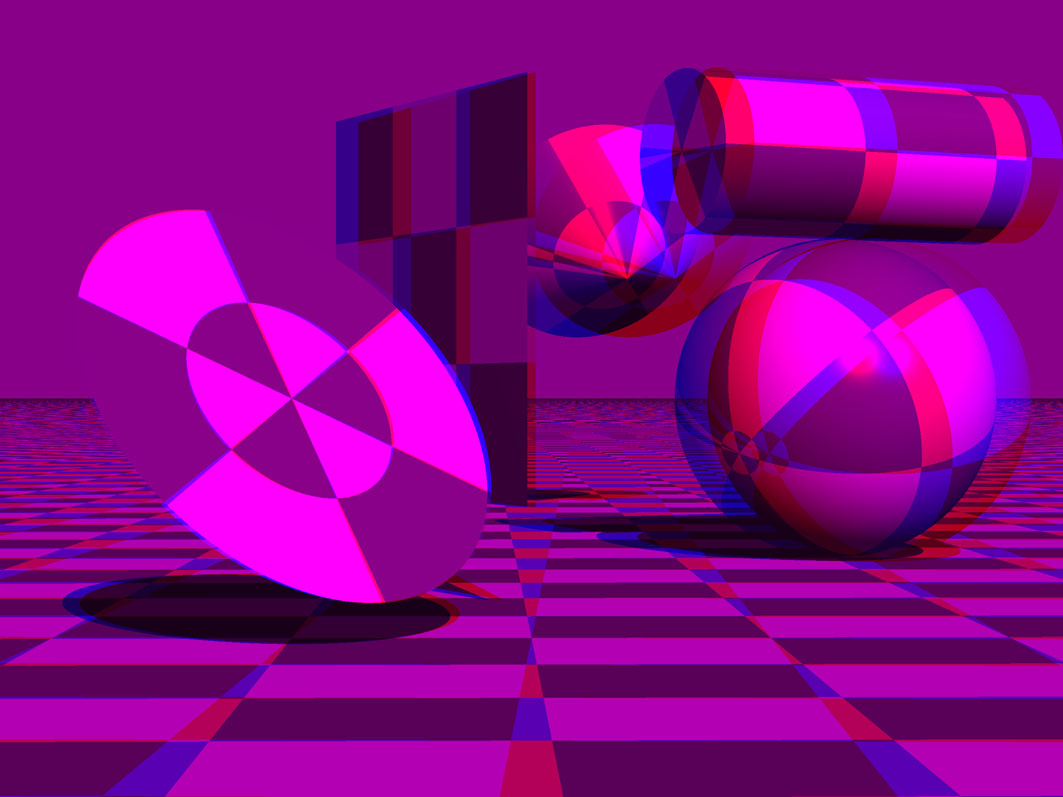} \end{center}
\caption{\label{tiled-objects-figure}Collection of scene objects, rendered in ``Eye view'' (top) and in ``Anaglyph 3D'' view.
}
\end{figure}

The view ``Anaglyph 3D'' creates an anaglyph image of the scene, designed for viewing with standard red/blue (or red/cyan) anaglyph glasses \cite{Wikipedia-AnaglyphImage}.
Anaglyph images fundamentally show two different views of the scene, colour-coded so that each eye sees one view.
In TIM, the left and right eyes see views from positions to the left and right of the ``Eye view'' position, $(0, 0, 0)$.
The ``Anaglyph view'' can also simulate relativistic camera movement through the scene (section \ref{relativistic-section}).
Fig.\ \ref{tiled-objects-figure} shows an example of the same scene rendered with the standard ``Eye view''  and ``Anaglyph 3D'' view.

\begin{figure}
\begin{center}
\includegraphics[width=\columnwidth]{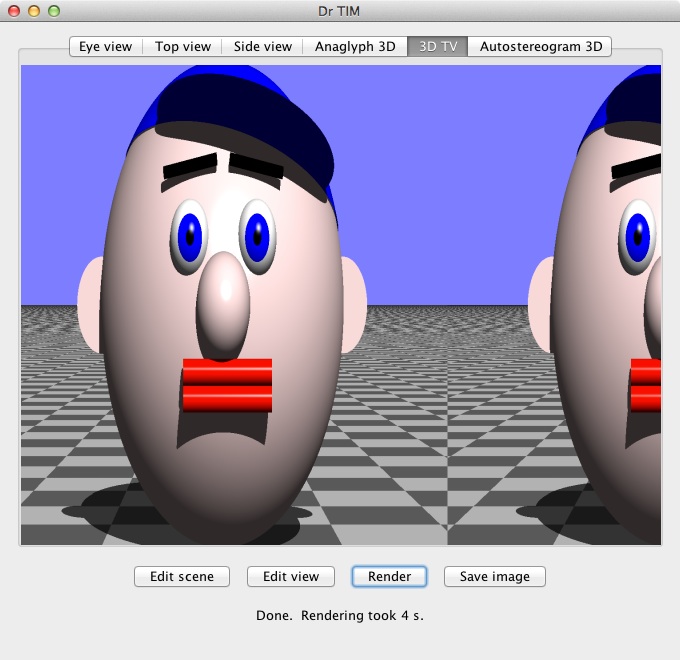}
\end{center}
\caption{\label{3DTV-figure}TIM's window after the default scene has been rendered in ``3D TV'' view using horizontal frame packing.
The left part of the image --- the left ``frame'' --- shows the image (with half the horizontal resolution) as seen from the left eye, the right part/frame (which is clipped when viewed in the Java application/applet, as shown here, but which is complete in the saved image resulting from pressing the ``Save image'' button) shows the image as seen from the right eye.
An HDMI 1.4a compatible 3D TV should be capable of displaying the saved image in 3D.}
\end{figure}

The ``3D TV'' view allows the creation of 3D image files in a number of HDMI-1.4a compatible formats.
All supported formats use frame packing (the image consists of two ``frames'', one containing the image as seen from the position of the left eye the other from the position of the right eye).
The 3D images produced in this view are not intended for viewing in TIM (in fact, often TIM shows a clipped image --- see Fig.\ \ref{3DTV-figure}), but for saving to file and display it on HDMI 1.4a compliant 3D TV.
The simulated 3D image can represent relativistic camera movement through the scene (section \ref{relativistic-section}).

\begin{figure}
\begin{center} \includegraphics[width=\columnwidth]{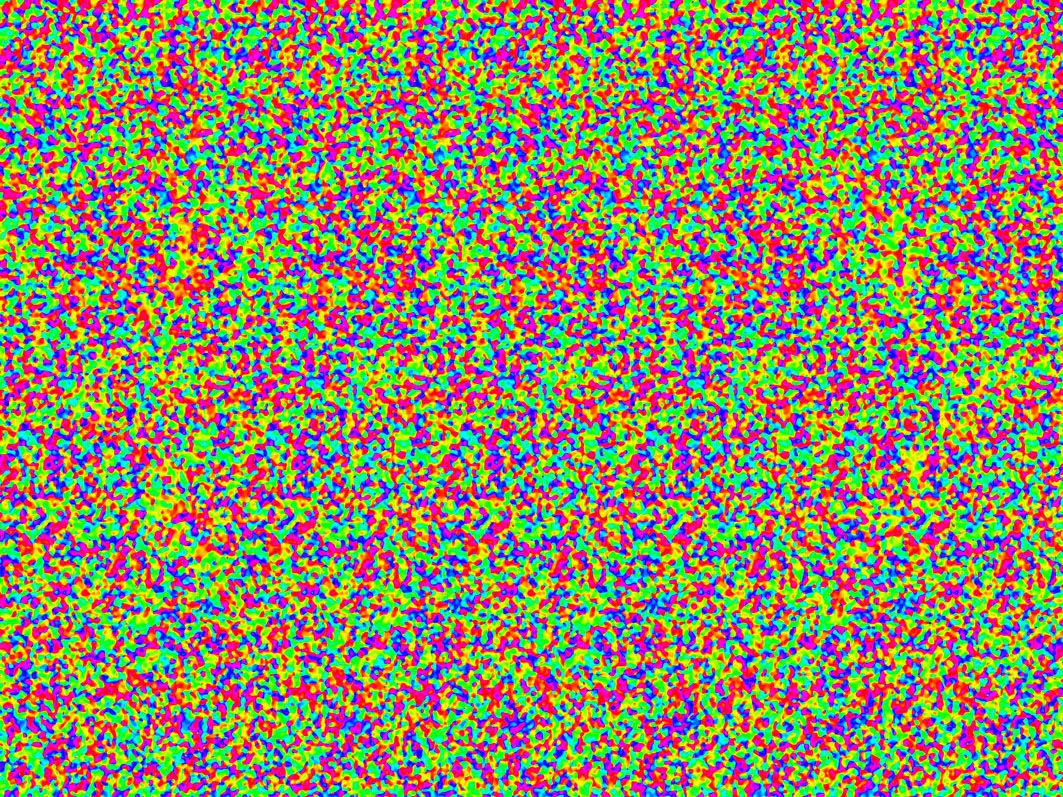} \end{center}
\caption{\label{autostereogram-example-figure}Random-dot autostereogram of the default scene shown in Fig.\ \ref{home-screen-figure}.
The plane behind Tim's head, which is invisible in the other views, makes it easier to perceive the scene's depth.}
\end{figure}

Finally, the view ``Autostereogram 3D'' creates a random-dot autostereogram (``Magic eye'') image \cite{Tyler-Clarke-1990} of the scene.
Details of how random-dot autostereograms work can be found elsewhere \cite{Lambert-et-al-2012}.
Note that autostereograms often work better for scenes that are not too complex and in which the depth range is quite limited.
For this latter reason the default scene contains a transparent plane (``invisible wall'') behind Tim's head that is invisible in other views (it is not only transparent, but also does not throw any shadow), but which is nevertheless visible in the ``Autostereogram 3D'' view as this ignores surface properties (see Fig.\ \ref{autostereogram-example-figure}).

The parameters of a specific view can be altered by clicking on the tab corresponding to that view and then clicking on the ``Edit view'' button.
All views allow the ``Anti-aliasing quality'' to be selected.
TIM implements anti-aliasing by calculating the image at a size that is greater than its screen size.
Before displaying the image on the screen (by default at size $640 \times 480$), the colour of the screen pixels is then calculated by averaging over a number pixels of the calculated image.
Table \ref{anti-aliasing-quality-table} lists the size at which the image is calculated for the different anti-aliasing qualities.
Note that anti-aliasing quality ``Normal'' calculates one pixel per screen pixel, and so calculates a sharp --- but aliased --- image.
The better anti-aliasing-quality settings calculate $2 \times 2$ (``Good'') or $4 \times 4$ (``Great'') image pixels per screen pixel, and respectively take 4 and 16 times longer to render.
There are also settings (``Bad'' and ``Rubbish'') which calculate the image at a \emph{reduced} size.
These are intended to provide previews.

``Anaglyph 3D'' allows editing of additional parameters:
``Centre of view'', which is the position where the centres of the field of view of the left eye and the right eye intersect;
``Eye separation'', which is a vector along the separation between the position of the left eye and the right eye;
and ``Colour'', which switches between different algorithms for calculating the anaglyph image, one that gives a colour impression, another that gives a monochromatic impression.

\begin{table}
\begin{center}
\begin{tabular}{c|c}
anti-aliasing quality & size of calculated image \\
\hline
Rubbish & $160 \times 120$ \\
Bad & $320 \times 240$ \\
Normal & $640 \times 480$ \\
Good & $1280 \times 960$ \\
Great & $2560 \times 1920$
\end{tabular}
\end{center}
\caption{\label{anti-aliasing-quality-table}Anti-aliasing quality descriptions and corresponding size at which the image gets calculated before being displayed.}
\end{table}

\section{\label{focussing-section}Focussing and aperture}

\noindent
In top view and side view, all objects are sharp (``in focus'').
These views can be understood as limiting cases of pinhole-camera views\footnote{As a pinhole camera moves further away, and the field of view becomes correspondingly smaller, it becomes ``more orthographic''.  In the limit of infinite camera distance and correspondingly zero field of view, the camera \emph{is} orthographic.}.

\begin{table}
\begin{center}
\begin{tabular}{c|c}
aperture size & radius \\
\hline
Pinhole & 0 \\
Small & 0.025 \\
Medium & 0.05 \\
Large & 0.1 \\
Huge  & 0.2
\end{tabular}
\end{center}
\caption{\label{aperture-size-table}Aperture-size settings and corresponding aperture radii.}
\end{table}

\begin{figure}[ht]
\begin{center} \includegraphics[width=\columnwidth]{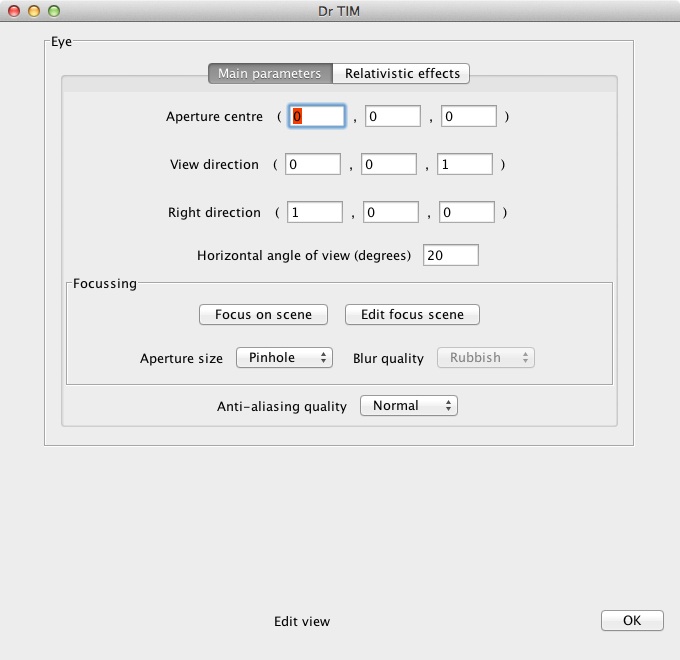} \end{center}
\caption{\label{edit-view-dialog-figure}Dialog for editing the ``Eye view''.
This can be accessed by selecting the ``Eye view'' tab and then clicking on the ``Edit view'' button.}
\end{figure}

\begin{figure}[ht]
\begin{center}
\includegraphics[width=8.4cm]{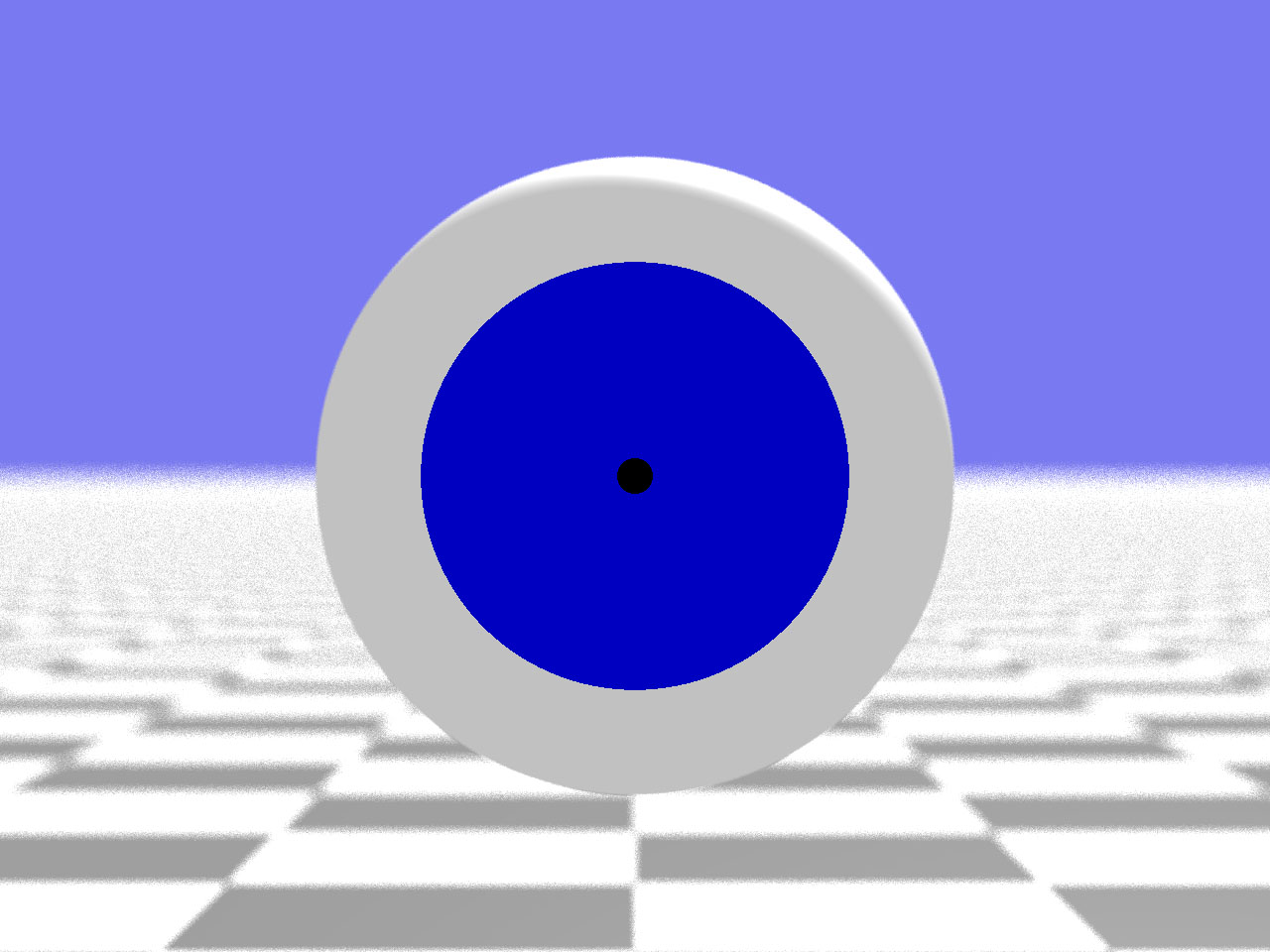} \\
\vspace{0.25cm}
\includegraphics[width=8.4cm]{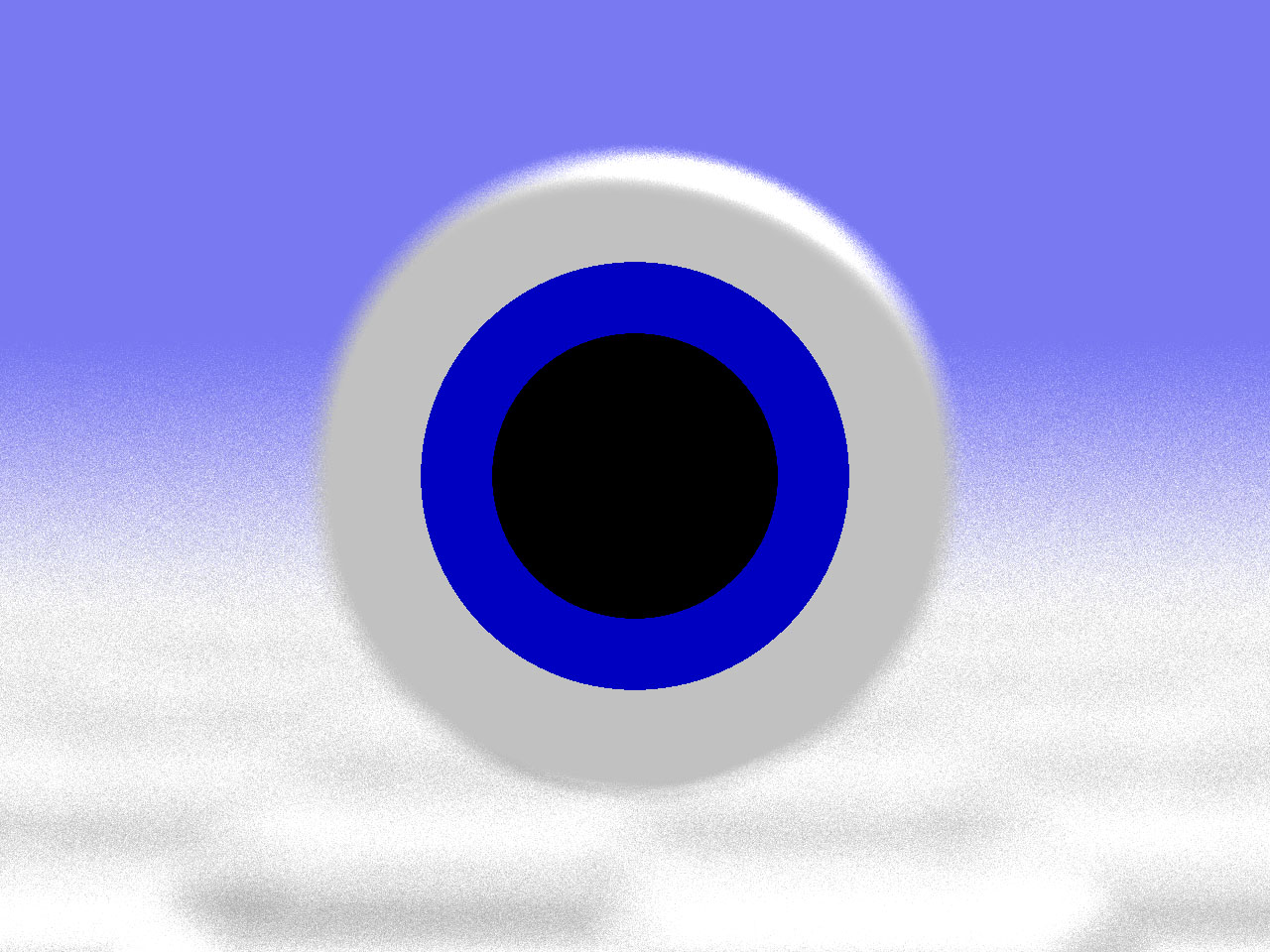}
\end{center}
\caption{\label{pupils-figure}The eye, seen in a mirror (a ``Rectangle'' scene object with surface type ``Reflective'') at $z=2$ with different aperture (pupil) sizes, namely ``Small'' (top) and ``Huge'' (bottom).
In both cases, the eye was focussed on a plane at $z=4$ so that the pupil plane, which is an optical path length of twice the eye-mirror distance from the eye, was in focus.
To increase the contrast between the eye's iris and pupil, the ambient brightness (see Ref.\ \cite{Lambert-et-al-2012}) was temporarily increased in the TIM source code for this calculation.}
\end{figure}

``Eye view'' also can also simulate a pinhole camera.
Unlike ``Top view'' and ``Side view'', it can additionally simulate a camera with a lens that has a circular aperture of finite size.
The aperture size can be adjusted in the dialog for changing the parameters for ``Eye view'' (Fig.\ \ref{edit-view-dialog-figure}).
Table \ref{aperture-size-table} lists the possible aperture sizes.
TIM can directly visualise the aperture size as the pupil size of a stylised eye (Fig.\ \ref{pupils-figure}).

\begin{table}
\begin{center}
\begin{tabular}{c|ccccc}
blur quality & rays per pixel \\\hline
Rubbish & 1 \\
Bad & 3 \\
Normal & 10 \\
Good & 32 \\
Great & 100
\end{tabular}
\end{center}
\caption{\label{rays-per-pixel-table}Blur-quality settings and corresponding number of rays used to calculate the colour of a single pixel.}
\end{table}

TIM simulates the effect of a finite-size aperture as follows.
Consider the calculation of the colour of light that hits a specific detector pixel.
The camera lens creates an image of the detector pixel (see section \ref{non-planar-focussing-section}), which means that any light ray that comes from the direction of the 3D position of this image and passes through the lens is redirected by the lens into the detector pixel.
When TIM traces a ray from the detector pixel backwards, it utilises the imaging property of the lens by tracing a ray that starts from a random position within the lens's aperture opening and that points in the direction of the 3D position of the pixel's image.
This is repeated for a number of rays starting at different aperture points, and the RGB brightnesses due to the different rays are averaged.

To understand how this leads to focussing or blurring, consider a scene that consists exclusively of coloured, non-reflective and non-transmissive, surfaces.
If the backwards-traced rays from one particular detector pixel all end on the same point on the surface of one of the scene objects --- and they only intersect in one point if this point is the image of the detector pixel 
--- then the averaged RGB brightness is the colour of this surface point;
if the rays do not all end on the same point, then the averaged RGB brightness will be the average of the colours of the various surface points where the different rays end.
The former case describes the creation of a sharp image of the intersection point; the latter case describes blurring.

Ideally, the contribution of \emph{every} point within the aperture opening is taken into account, but this requires tracing of infinitely many rays.
In practice, a finite number of rays are used; the greater the number of rays, the higher the quality of the blur in the rendered image, which is why the number of rays per pixel can be adjusted by altering the ``Blur quality'' setting in the ``Eye view'' dialog.
Table \ref{rays-per-pixel-table} lists all possible ``Blur quality'' settings and the corresponding number of rays per pixel.

\section{\label{non-planar-focussing-section}Focussing on non-planar surfaces}

\noindent
One of TIM's unique features is the ability to focus on non-planar surfaces.
We call the surface onto which a camera is focussed its \emph{focus surface}.

The camera lens forms an image of each detector pixel, and the focus surface is determined by the positions of the images of \emph{all} detector pixels (section \ref{focussing-section}).
With a standard camera lens, because the detector pixels lie in a plane perpendicular to the lens's optical axis, the images of all these pixels also lie in a plane\footnote{Strictly speaking, the images of the detector pixels lie in a plane only approximately, due to lens aberrations \cite{Wikipedia-PetzvalFieldCurvature}.} perpendicular to the lens's optical axis.
This can be generalised with a perspective-control lens (also called tilt-shift lens) \cite{Wikipedia-PerspectiveControlLens}, in which the lens --- and with it its optical axis --- can be tilted with respect to the plane of the detector, and so the plane of the image of the detector is also tilted with respect to the lens's optical axis.

TIM can focus on considerably more general surfaces.
This works as follows.
In section \ref{focussing-section} we discussed how blurring is related to imaging by the camera lens.
Specifically, the RGB brightness recorded by a specific detector pixel is calculated as the average of the RGB brightnesses of a number of rays that hit this pixel.
This involves tracing rays backwards, starting from different positions on the lens aperture and travelling in the direction of the 3D position of the image of the detector pixel.
TIM uses this procedure, but with a twist:  the 3D position of the image of a detector pixel is not determined by standard imaging with a lens.
Instead, TIM defines an additional collection of scene objects, the ``focus scene'';
establishes the 3D position on the surface of the focus-scene object that a backward-traced ray would first intersect if the camera was a pinhole camera;
and then uses this position as the position of the image of the detector pixel.
If the focus scene contains only a plane $z = z_0$, then focussing is the same as with a traditional lens that is focussed at distance $z_0$.
Note that TIM is not concerned with how an actual lens (or other imaging element) would achieve this, if indeed it can be achieved at all optically.
It should be possible to achieve it using computational imaging with a plenoptic camera (or light-field camera)~\cite{Adelson-Wang-1992,Ng-et-al-2005}.

\begin{figure}
\begin{center}
\includegraphics[width=8.4cm]{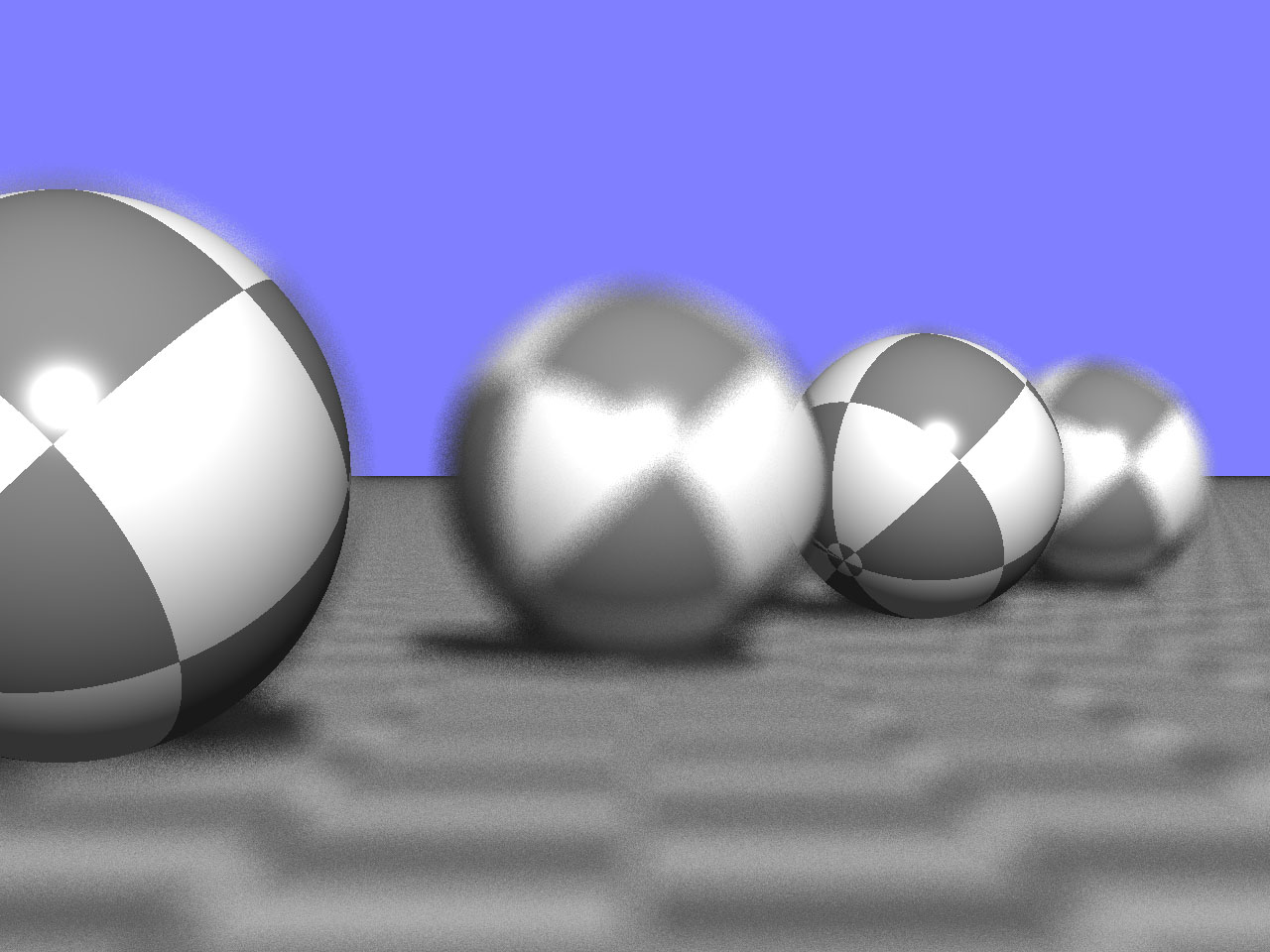} \\
\vspace{0.25cm}
\includegraphics[width=8.4cm]{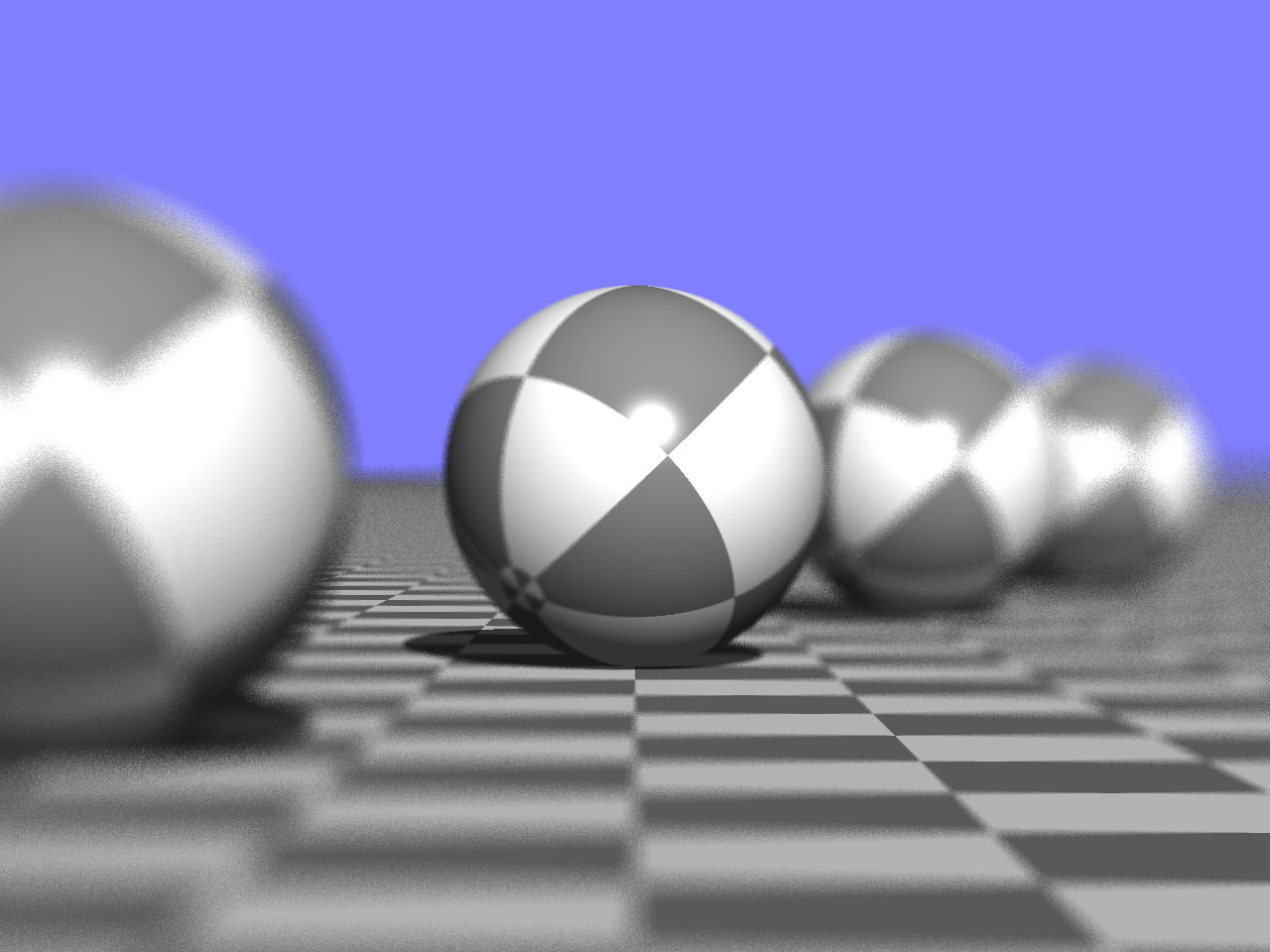}
\end{center}
\caption{\label{four-spheres-figure}Focussing on different surfaces.
The scene consists of four spheres located at different distances from the camera.
In the top image, the focus scene contains two of these spheres and the sky;
in the bottom image it consists of a plane that passes through the centre of one of the spheres and that is not perpendicular to the $z$ axis (the floor tiles are in focus along the line where this plane intersects the floor).
The spheres have unit radius; they are centred respectively at $(-2, 0, 10)$, $(0, 0, 15)$, $(2, 0, 20)$, and $(4, 0, 25)$.
All four spheres have identical tiled surfaces (edit sphere; select surface type ``Tiled''; in the ``Parametrisation'' tab (see Fig.\ \ref{sphere-parametrisation-panel-figure}), set ``Direction from centre to north pole'' to $(1, 1, 1)$).
The focus plane in the bottom frame is described by the point on the plane $(0, 0, 15)$ and the normal to the plane $(4, 0, 1)$ (so the plane is vertical and has an angle of $14^\circ$ with respect to the $z$ direction).
The image was calculated with ``Large'' aperture size and ``Good'' blur quality and anti-aliasing quality.}
\end{figure}

The following example illustrates how to render scenes focussed on non-planar surfaces.
We assume that we have already defined a scene consisting of four spheres at different $z$ distances from the camera.
For the focus scene to have any effect we need a finite-size aperture; this can be achieved by selecting any aperture size other than ``Pinhole'', for example ``Large'' (see section \ref{focussing-section}).
We can now focus on two of the spheres and the sky\footnote{Note that TIM needs to be able to find an image position for every detector pixel, so we need to define a focus-scene object for those pixels whose image does not lie on one of the two in-focus spheres.
This is easily achieved by including the sky --- a large sphere completely surrounding the camera --- in the focus scene.} as follows.
First we copy the scene into the focus scene; this is done by clicking on the ``Focus on scene'' button in the dialog for editing the ``Eye view'' parameters.
Then we remove all objects we do not want to focus on from the focus scene; we do this by clicking on the ``Edit focus scene'' button and either removing them or making them invisible (by unchecking the ``Visible'' checkbox).
Clicking on ``OK'' twice takes us back to TIM's main screen, from where we can now render the image (by clicking on the ``Render'' button), which is shown in the top frame of Fig.\ \ref{four-spheres-figure}.

It is perhaps worth noting that the focus scene can be edited just like the scene itself.
Specifically, we can add objects to the focus scene, for example to focus on an inclined plane, like a camera with a tilt-shift lens (bottom frame of Fig.\ \ref{four-spheres-figure}).
Only scene objects for whom ``Visible'' is checked are part of the focus scene.
TIM ignores the surface properties of focus-scene objects.

\section{\label{relativistic-section}Relativistic camera speeds}

\begin{figure}[ht]
\begin{center}
\includegraphics[width=\columnwidth]{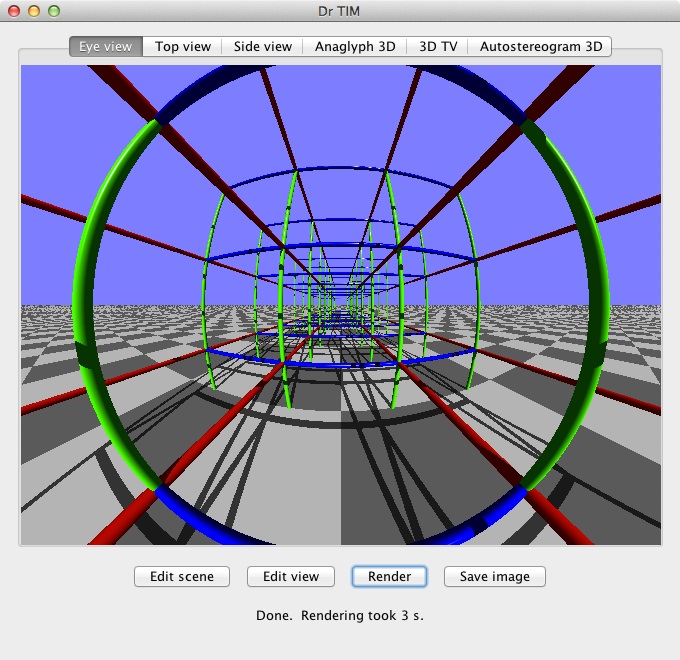}
\end{center}
\caption{\label{relativistic-lattice-figure}TIM's window after simulating a camera moving at relativistic speed.
The scene is the cylinder lattice (in addition to the floor and sky) that results from selecting ``Relativistic (large lattice)'' in the ``Initialise scene to...'' drop-down menu in the ``Edit scene'' dialog.
The view is the standard ``Eye view'', but with the relativistic parameters as shown in Fig.\ \ref{relativistic-parameters-figure}.}
\end{figure}

\noindent
Dr TIM is capable of simulating snapshot taken with a camera that is moving at relativistic speed through a scene of stationary objects; Fig.\ \ref{relativistic-lattice-figure} shows an example.
The theory behind this new capability is explained in some detail in Ref.\ \cite{Oxburgh-et-al-2013}.
The moment in simulated time when the snapshot is taken can be varied, allowing the creation of movie frames.
This new capability can be combined with several of TIM's other capabilities, such as simulating a camera with a finite-size aperture and displaying 3D images.

TIM simulates a camera moving, with constant velocity $\bm{v}$, through a scene of stationary objects.
TIM's algorithm only considers the change in the position in which objects are seen, an effect known as relativistic aberration.
It neglects all other effects, specifically the Doppler effect (which alters colour non-isotropically) and the headlight effect (which alters the brightness non-isotropically)~\cite{Savage-et-al-2007}.

\begin{figure}[ht]
\begin{center}
\includegraphics[width=\columnwidth]{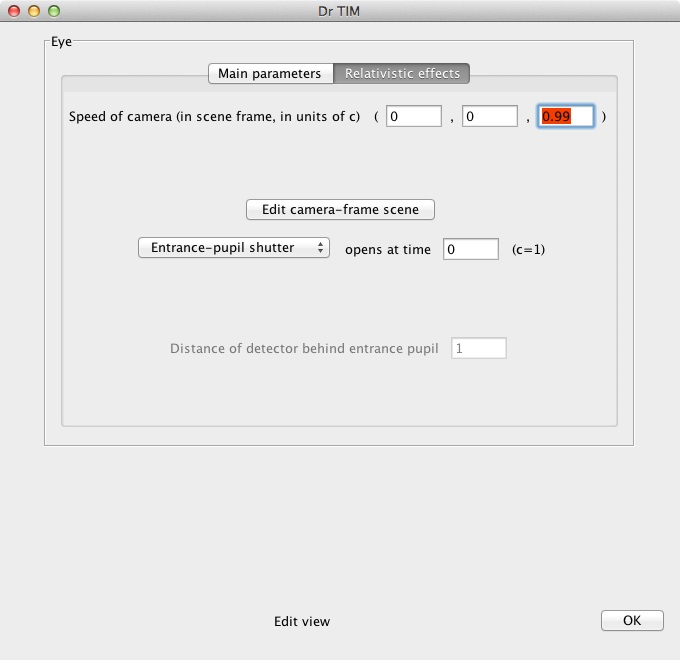}
\end{center}
\caption{\label{relativistic-parameters-figure}Panel for editing the parameters controlling relativistic effects of the ``Eye view''.
Here, the $z$ component of the speed of the camera (i.e.\ the component in the view direction) is set to 99\% of the speed of light, $c$.}
\end{figure}

\begin{figure}[ht]
\begin{center}
\includegraphics{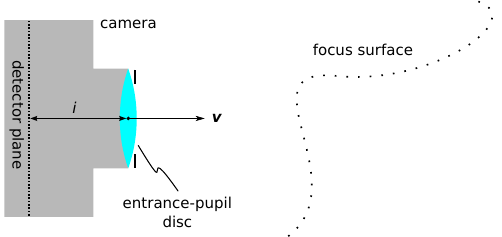}
\end{center}
\caption{\label{camera-figure}TIM's shutter surfaces.}
\end{figure}

The time the ray leaves the camera frame affects the position of the same event in the scene frame.
This time is determined by the shutter model used.
Every camera has some form of shutter somewhere within its body.
In modern SLRs, for example, the shutter is usually in a plane immediately in front of the detector plane.
TIM can generalise this idea to other shutter surfaces (Fig.\ \ref{camera-figure}):
\begin{enumerate}
\item the \emph{detector plane}, a plane perpendicular to the view direction $\bm{v}$ a distance $i$ behind the entrance pupil;
\item the \emph{entrance-pupil disk}, positioned immediately in front of or behind the lens or generalised focussing element (so that TIM can focus on arbitrary surfaces --- see section \ref{focussing-section});
\item the \emph{focus surface}, the surface which is imaged to the detector plane (section~\ref{focussing-section}).
\end{enumerate}
In TIM's shutter model, the shutter is open for an instant at time $t_s$.
In the camera frame, it opens simultaneously across the entire shutter surface.

\section{\label{light-rays-section}Visualising light-ray trajectories}

\noindent
TIM has the capability to visualise light-ray trajectories.
This is implemented by first tracing those light rays through the scene, recording the start and end points of each straight-line trajectory segment;
adding to the scene a cylinder along every one of these straight-line segments;
and finally rendering the extended scene.

\begin{figure}
\begin{center}
\includegraphics[width=8.4cm]{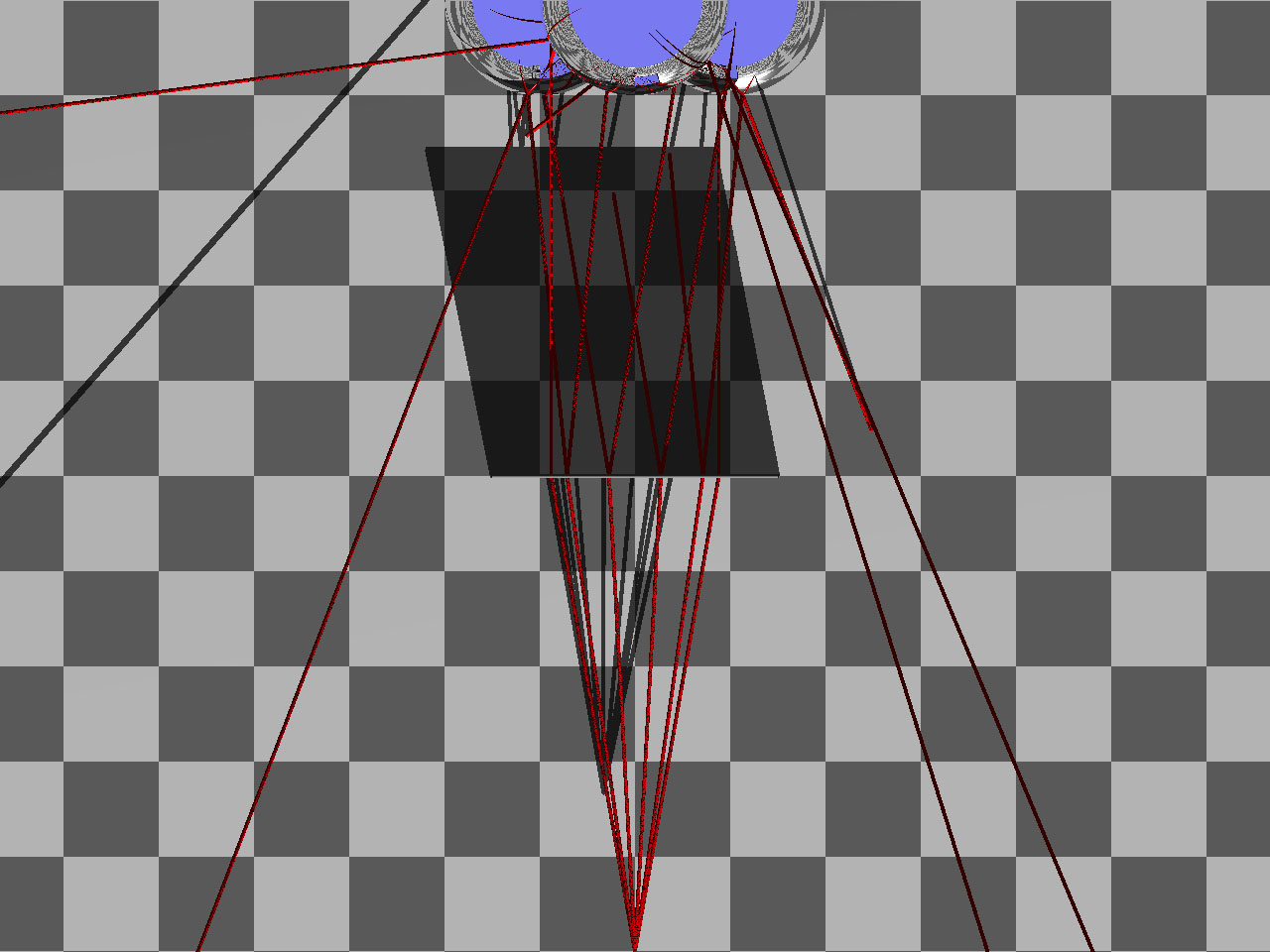} \\
\vspace{0.25cm}
\includegraphics[width=8.4cm]{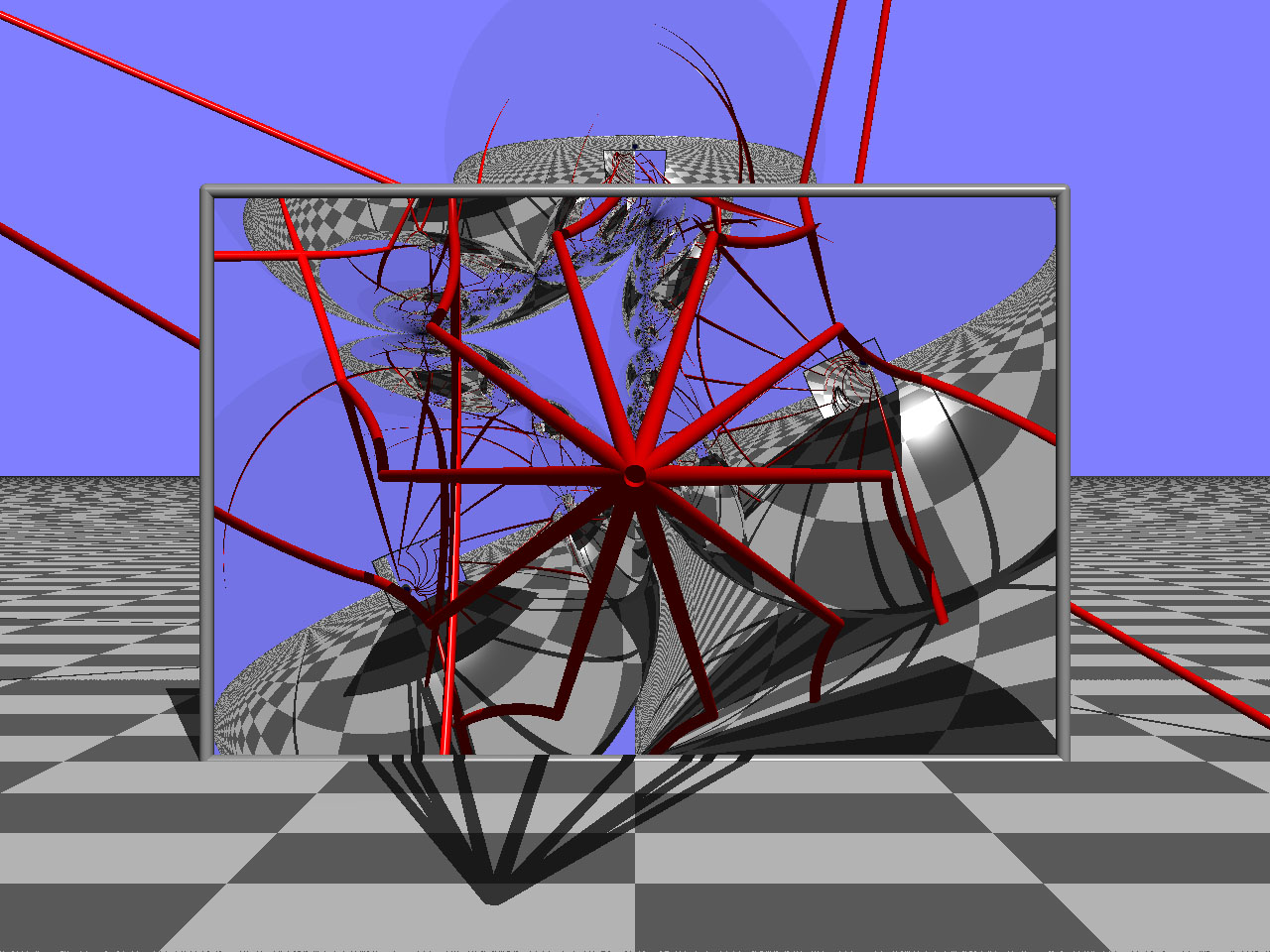}
\end{center}
\caption{\label{light-ray-trajectories-figure}Visualisation of the trajectories of a cone of light rays through a complex scene that includes a ray-rotating METATOY window and reflecting spheres.
(Top)~Top view; (bottom)~eye view.
The frames were calculated with anti-aliasing quality ``Good''.}
\end{figure}

Fig.\ \ref{light-ray-trajectories-figure} shows the trajectories of a cone of light rays launched into one of TIM's standard scenes, which can be chosen by selecting, in the ``Edit scene'' dialog's ``Initialise scene to...'' drop-down menu (see Fig.\ \ref{edit-scene-dialog-figure}), ``Original (shiny spheres behind ray-rotating window)''.
The cone of light rays (``Ray-trajectory cone'') was simply added to the scene with all parameters taking default values, apart from the start point, which is $(0, 0, 5)$, and the cone angle, which is $10^\circ$.
Clicking on the ``Render'' button on the main screen then renders the scene and the light-ray trajectories.
Note that this works not only in ``Eye view'', but also in the other views (see top frame of Fig.\ \ref{light-ray-trajectories-figure}).
Note also that segments of light-ray trajectories that are seen through METATOYs, reflected off curved mirrors, etc., are distorted accordingly.
This can clearly be seen in the bottom frame of Fig.\ \ref{light-ray-trajectories-figure}, in which many of the straight-line trajectory segments appear bent when seen through the ray-rotating METATOY.

\section{\label{parametrisation-section}Parametrisation of scene objects}

\noindent
Scene objects in TIM all have associated with them coordinate systems that describe their surface.
Each point on the surface of a scene object is described by the corresponding values of these coordinates, the point's \emph{surface coordinates}.
The surface-coordinate systems were chosen to be the most ``natural'' choice; for example, the surface of a sphere is parametrised in terms of spherical polar coordinates \cite{Wikipedia-SphericalCoordinateSystem}, i.e.\ the polar angle $\theta$ and the azimuthal angle $\phi$.

\begin{figure}
\begin{center} \includegraphics{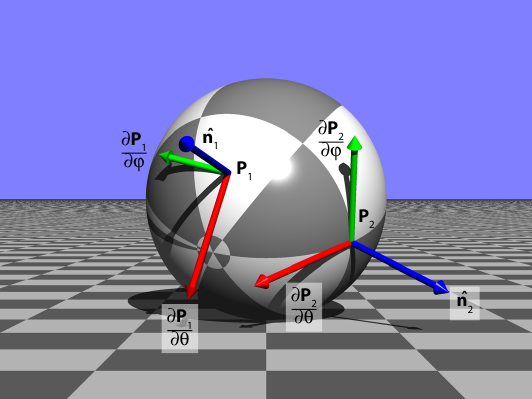} \end{center}
\caption{\label{parametrisation-figure}Parametrisation of a scene object.
Each point on the surface is described by a pair of surface coordinates, in the picture the spherical coordinates $\theta$ and $\phi$, defined with respect to an arbitrary zenith direction (here $(0.408, 0.408, 0.816)$) and direction of the azimuth axis, i.e.\ the direction from the centre to the $0^\circ$ meridian (here $(0.913, -0.183, -0.365)$).
This parametrisation of the surface has been indicated by covering it in a chequerboard pattern with tiles of side length 1 in both $\theta$ and $\phi$.
The local surface-coordinate axes, $\hat{\mathbf{\theta}}_i = \partial \textbf{P}_i / \partial \theta$ and $\hat{\mathbf{\phi}}_i = \partial \textbf{P}_i / \partial \phi$, together with the local surface normals, $\hat{\mathbf{n}}_i$, are shown for two points, $\mathbf{P}_i$ ($i = 1, 2$).
The sphere has radius 1 and is centred at $(0, 0, 10)$.}
\end{figure}

Instead of explaining in detail how surfaces are parametrised, we have built into TIM a number of features that encourage interactive exploration of this parametrisation once a scene has been rendered:
\begin{itemize}
\item hovering with the mouse cursor over any point of the rendered image displays the global coordinates (section \ref{views-section}) of the point $P$ on the surface of the scene object the corresponding backwards-traced light ray first intersects;
\item clicking on any point of the rendered image displays the surface coordinates of $P$;
\item right-clicking brings up a popup menu that allows the addition of the axes of the local surface-coordinate system at $P$; the scene then needs to be re-rendered for the axes to be shown.
\end{itemize}
Fig.\ \ref{parametrisation-figure} shows an example of a scene object with sets of surface-coordinate-system axes added at two points on its surface.

The parametrisation of surfaces can also be glimpsed by rendering scene objects with a ``Tiled'' surface.
The sphere in Fig.\ \ref{parametrisation-figure} is one example; several other examples are scattered throughout this document, the most useful of these is perhaps Fig.\ \ref{tiled-objects-figure}.

\begin{figure}
\begin{center}
\includegraphics[width=\columnwidth]{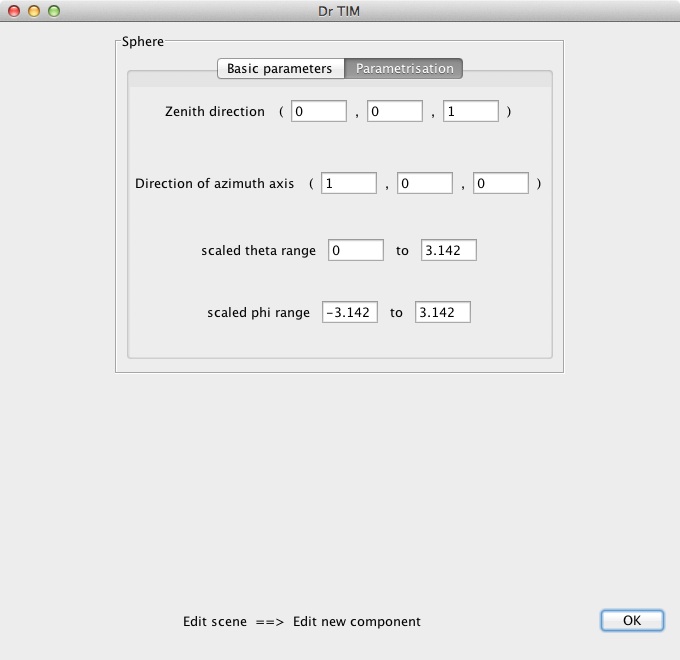}
\end{center}
\caption{\label{sphere-parametrisation-panel-figure}Panel controlling the parametrisation of a (newly created) sphere.}
\end{figure}

Most scene objects allow the parametrisation of their surface to be controlled.
Fig.\ \ref{sphere-parametrisation-panel-figure}, for example, shows the panel that controls the parametrisation of a sphere.
This panel is accessed by clicking on the ``Parametrisation'' tab when editing a sphere, which is parametrised in terms of the polar angle $\theta$ and the azimuthal angle $\phi$.
It allows the directions of different axes to be altered, specifically the ``Zenith direction'', which corresponds to the direction for which $\theta=0$, and the ``Direction of azimuth axis'', where $\phi=0$.
It is worth noting that clicking on ``OK'' performs a few checks on the entered parameters and corrects them, if necessary.
Specifically, to ensure that the direction of the azimuth axis is perpendicular to the zenith direction, the vector made up of the entered component values is projected into the plane perpendicular to the zenith direction.
Furthermore, both vectors are normalised.
It is also worth noting that the range of both coordinates can be scaled.

Additional details on coordinates can be found in Ref.\ \cite{Lambert-et-al-2012}.

\section{Conclusions}

\noindent
We hope you find TIM useful, (reasonably) user-friendly, and fun.
Perhaps you even feel a warm glow in your heart whenever you set eyes on those luscious lips.

\bibliography{/Users/johannes/Documents/work/library/Johannes}

\end{document}